\documentclass[11pt,a4paper]{article} 

\usepackage{jheppub}
\usepackage{graphicx}

\usepackage{amsmath}
\usepackage{amssymb}
\usepackage{accents}
\usepackage{hyperref}

\def\dual#1{\accentset{\boldsymbol{\neg}\vspace{-0.2ex}}{#1}}

\title{Lorentz invariant dark-spinor and inflation}
\author[\dagger]{Abhishek Basak}
\author[\ddagger]{and Jitesh R. Bhatt}
\affiliation[]
{Theoretical Physics Division, Physical Research Laboratory,\\
Navarangpura, Ahmedabad, India}
\emailAdd{ abhishek@prl.res.in}
\emailAdd{jeet@prl.res.in}

\abstract{We investigate the possibility of the inflation driven by
a Lorentz invariant non-standard spinor field. As these spinors are having
dominant interaction via gravitational field only, they are 
considered as \emph{Dark Spinors}. We study how these dark-spinors can drive
the inflation and investigate the cosmological (scalar) perturbations generated
by them. Though the dark-spinors obey a Klein-Gordon like equation, 
the underlying theory of the cosmological perturbations
is far more complex than the theories which are using a canonical scalar
field. For example the sound speed of the perturbations
is not a constant but varies with time. We find that in order to explain 
the observed value of the spectral-index $n_s$ one must have upper bound
on the values of the background NSS-field. The tensor to scalar ratio remains as small as that in 
the case of canonical scalar field driven inflation because the correction to tensor 
spectrum due to NSS is required to be very small. In addition we discuss the relationship
of results with previous results obtained by using the Lorentz invariance
violating theories.}

\keywords{dark spinors, non-standard spinors, inflation, cosmological-perturbations, Lorentz invariance}

\begin{document}
\maketitle
\section{Introduction}
High precision cosmology has significantly changed our idea of the
universe. The present state of the universe can be explained very well
with the assumption that very early during its history the universe
has undergone an inflationary phase of the expansion \cite{guth,riotto,mukh,wein}.
In addition the measurements of Type-I supernovae around redshift $z\sim 1$,
together with the other measurements, suggest that the universe is currently
undergoing an accelerated phase of expansion for the second time in its history
after the big bang. This phase of the accelerated expansion is attributed
to dark-energy \cite{dewhite,pady}. According to our present understanding the
total energy of the universe is very close to the critical energy and
the baryonic matter constitute only  4$\%$ of it. While 
the dark-matter and dark-energy contribute  around 22$\%$ and 74$\%$ of the total
energy respectively \cite{desah}. Presence of the dark-sector in 
the energy budget 
may be indicative of the fact that either general relativity or the standard model of particles 
(or both) are inadequate to explain the current astrophysical and cosmological data.
Thus in such a situation it is worthwhile to look 
for new kind of particles or fields that can be 
candidates for the dark-matter, dark-energy and inflation. 

Recently there is a lot of interests in studying dark or Non-Standard Spinor (NSS). 
The theory of NSS was first developed in Refs.\cite{ah1,ah2}. 
Subsequently the NSS models were further developed and investigated by several authors
\cite{rocha1,rocha2,ah3,ah4,luca,luca1,luca2}.
These spinors can  be regarded as `dark' as their dominant interaction is
via gravitational field only and they have been extensively applied to study above mentioned
problems in cosmology \cite{bh,sh1,sh2,sh3,bh1,bh2,bh3,bh4,hwei,myrza1,myrza2}.
Unlike the Dirac, Majorana or Weyl spinors, NSS propagator behaves like $1/p^2$
in the large momentum limit and has mass dimension one. At present the theory
of NSS is under development, however, NSS are known to be  either violating the Lorentz invariance
or locality or both. Basic Lagrangian of NSS can be written as
\begin{equation}
\mathcal{L}_{\rm cosmo} = \frac{1}{2} \dual{\lambda} \overleftarrow{\nabla}_{\mu} \nabla^{\mu}\lambda - 
V(\dual{\lambda}\lambda),
\end{equation}
where, $\dual{\lambda}\overleftarrow{\nabla}_{\mu}\equiv \partial_{\mu}\dual{\lambda} + \
\dual{\lambda}\Gamma_{\mu} $, 
$\nabla_{\mu} \lambda \equiv \partial_{\mu} \lambda - \Gamma_{\mu} \lambda$. 
$\lambda$ and $\dual{\lambda}$ are NSS and its dual respectively.
$\Gamma_\mu$ are defined as
\begin{eqnarray}
\label{eq:spincon}
\Gamma_{\mu} = \frac{i}{4}\omega^{ab}_{\mu} f_{ab}, \qquad
f^{ab} = \frac{i}{4}\left[ \gamma^{a}, \gamma^{b}\right],
\end{eqnarray}
where index $\mu$ is the space-time index and index $a$ is the spinor index.
Here $\omega^{ab}_{\mu}$ is defined as 
\begin{equation*}
 \omega_{\mu}^{ab} = e_{\nu}^{a}\partial_{\mu}e^{\nu b} + 
                                     e_{\nu}^{a}e^{\sigma b} \Gamma_{\mu \sigma}^{\nu},
\end{equation*}
where $e^{a}_{\mu}$ are tetrads defined as $e_{\mu}^{a}e_{\nu}^{b}\eta_{ab} = g_{\mu \nu}$. Here 
$g_{\mu \nu}=a^{2}{\rm diag}(1,-1,-1,-1)$ is the space-time metric, $a$ is the scale factor,
$\eta_{ab} = {\rm diag}(1,-1,-1,-1)$ and 
$\Gamma_{\mu \sigma}^{\nu}$ 
are Christoffel symbols of $g_{\mu\nu}$. $\gamma$-matrices are defined as
\begin{eqnarray*}
 \gamma^{0} = \begin{pmatrix} 0 & \mathbb{I}_{2\times 2} \\ \mathbb{I}_{2\times 2} & 0 \end{pmatrix}, \qquad
\gamma^{i} = \begin{pmatrix} 0 & -\sigma^{i} \\
 \sigma^{i} & 0 \end{pmatrix},
\end{eqnarray*}
where $\sigma^{i}$ $(i=1,2,3)$ are Pauli matrices defined as 
\begin{equation*}
 \sigma^{1} = \begin{pmatrix} 0 & 1 \\ 1 & 0 \end{pmatrix}, \qquad
\sigma^{2} = \begin{pmatrix} 0 & -i \\ i & 0 \end{pmatrix}, \qquad
\sigma^{3} = \begin{pmatrix} 1 & 0 \\ 0 & -1 \end{pmatrix}.
\end{equation*}

 It should be noted that the NSS theory described by the above Lagrangian is not
Lorentz invariant \cite{bh}. 
It has has been proposed in \cite{bh} that
the locally Lorentz invariant energy-momentum tensor $T^{\mu \nu}_{\rm cosmo}$ can be constructed from
$\mathcal{L}_{\rm cosmo}$ as
\begin{equation}
T^{\mu \nu}_{\rm cosmo} = \dual{\lambda} \overleftarrow{\nabla}^{(\mu}\nabla^{\nu)}\lambda - 
g^{\mu \nu} \mathcal{L}_{\rm cosmo} + \frac{1}{2}\nabla_{\rho} J^{\mu \nu \rho}, 
\end{equation}
where $J^{\mu \nu \rho}$ defined as
\begin{equation}
J^{\mu \nu \rho} = -i\left[\dual{\lambda} \overleftarrow{\nabla}^{(\mu}f^{\nu)\rho}\lambda + 
\dual{\lambda} f^{\rho (\mu} \nabla^{\nu)}\lambda\right]
\end{equation}
is the additional term which did not appear in the earlier non-Lorentz invariant models \cite{bh1,bh2,bh3,bh4}.
It can be argued that any reasonable theory of cosmological perturbations should be based on Lorentz
invariant frame work. Keeping this view in mind we study the characteristics of the inflation driven by NSS
given in \cite{bh}. As it turned out , the cosmological perturbations based on equation(1.3) are far more 
complex than the theory based upon canonical scalar field model. Appearance of $J^{\mu\nu\rho}$ term 
can give rise to an additional scale $\tilde{F}=\frac{\dual{\lambda}\lambda}{8M^{2}_{\rm pl}}$ in the problem,
where $M_{\rm pl}=\sqrt{\frac{1}{8\pi G}}$ is the reduced Planck mass (G is the gravitational constant).

It is generally assumed that
the inflation is driven by a scalar field, which can have the following verifiable predictions: (a)
nearly a scale invariant spectrum, (b) existence of gravitational waves and (c) the tensor 
to scalar power spectrum may be of the order of $\epsilon$, where $\epsilon$ is the slow roll 
parameter \cite{riotto,mukh,wein}.
At this juncture in order to gain further insight it is natural to question the elementary scalar field
inflation scenario. Therefore, in this work, we investigate some of the predictions of the inflation theory by
assuming that the inflation is driven by a NSS with energy-momentum tensor described by equation (1.3). 
We would like to note that a similar kind of study exist in the 
literatures but it is based on non-Lorentz invariant NSS \cite{sh1}.

In this study we are interested in calculating power spectrum of the scalar perturbations generated
by NSS and compare our results with the spectrum calculated from canonical scalar inflaton field. In section 2
we write the basic equations for homogeneous and isotropic background. In section 3 we give the details 
of the cosmological perturbations using $T^{\mu\nu}$ given in equation (1.3). While section 4 contains 
the calculation of power spectrum and its comparison with CMB spectrum.

\section{Unperturbed equations}
Here we consider the following structure of NSS ($\lambda$) and its dual ($\dual{\lambda}$):
\begin{equation}
\lambda=\varphi(\eta)\xi,   \qquad
 \dual{\lambda}=\varphi(\eta)\dual{\xi},
\end{equation}
where, $\varphi(\eta)$ is a scalar quantity and $\eta$ is the conformal time defined as 
$d\eta=\frac{dt}{a}$. 
$\xi$ and $\dual{\xi}$ are two constant matrices with 
$\dual{\xi}\xi=\mathbb{I}$. In a flat, isotropic and homogeneous space-time unperturbed 
$\mathcal{L}_{\rm cosmo}$ 
can be written as 
\begin{equation}
 \mathcal{L}_{\rm cosmo}=\frac{1}{2a^{2}}\left[ \varphi^{\prime 2}+
 \frac{3}{4}H^{2}\varphi^{2} \right]-V(\varphi^{2}),
\end{equation}
where prime ($^{\prime}$) denotes the derivative with respect to conformal time $\eta$ . 
$V (\varphi)$ is the potential which is a function of $\varphi$. While the Hubble expansion parameter $H$
is defined as $H=\frac{a^{\prime}}{a}$.
Unperturbed energy momentum tensors and equation of motion for $\varphi$
in FRLW space-time have already been 
calculated in \cite{bh}. 
We are enlisting them below in conformal time . 
Let us first define the covariant energy momentum tensor 
$T^{\mu\nu}_{\rm cosmo}$, which appears into the Einstein's equation, as 
\begin{equation} 
T^{\mu\nu}_{\rm cosmo}=\bar{T}^{\mu\nu}+\frac{1}{2}\nabla_{\rho}J^{\mu\nu\rho},
\end{equation}
where, 
\begin{equation*}
\bar{T}^{\mu\nu}= \dual{\lambda} \overleftarrow{\nabla}^{(\mu}\nabla^{\nu)}\lambda - g^{\mu\nu}\mathcal{L}_{cosmo}.
\end{equation*}
The non-vanishing components of $J^{\mu\nu\rho}$ are
$$
J^{i\eta j}=J^{\eta ij}=\frac{1}{4}\frac{H}{a^{4}}\varphi^{2}\delta_{ij}, \qquad
J^{ij\eta}=-\frac{1}{2}\frac{H}{a^{4}}\varphi^{2}\delta_{ij}. 
$$
Here $\varphi^{2}=\dual{\lambda}\lambda$, is a function of time only. 
Next, we can write the expressions for the energy density $\varepsilon$ and pressure $p$ as following:
\begin{eqnarray*}
\varepsilon=T^{\eta}_{\eta}=\bar{T}^{\eta}_{\eta}+F^{\eta}_{\eta}, \qquad
 p =-T^{i}_{j}\delta_{ij}=-\left(\bar{T^{i}_{j}}+F^{i}_{j}\right)\delta_{ij}.
\end{eqnarray*}
Expressions for $\bar{T}^{\mu\nu}$ and $F^{\mu\nu}$ can be written as,
\begin{equation*}
 \bar{T}^{\eta}_{\eta}=\frac{1}{2a^{2}}\left[\varphi^{\prime2}-\frac{3}{4}H^{2}\varphi^{2}\right]+V
\end{equation*}
and 
\begin{equation*}
F^{\eta}_{\eta}=\frac{3}{4a^{2}}H^{2}\varphi^{2}. 
\end{equation*}
From these one can write energy density as
\begin{equation}
 \varepsilon=\frac{1}{2a^{2}}\left[\varphi^{\prime2}+\frac{3}{4}H^{2}\varphi^{2}\right]+V.
\end{equation}
It is useful to write the expression for $\varepsilon$ as,
\begin{equation*}
 \varepsilon=X+V,
\end{equation*}
where, $X=\left(\nabla^{\eta} \dual{\lambda}\nabla_{\eta}\lambda\right)-
                 g^{\eta}_{\,\,\eta}\left(\frac{1}{2}\nabla_{\mu}\dual{\lambda}\nabla^{\mu}\lambda\right)+
                 g_{\eta\eta}\frac{1}{2}\nabla_{\rho}J^{\eta\eta\rho}
              =\frac{1}{2a^{2}}\left[\varphi^{\prime2}+\frac{3}{4}H^{2}\varphi^{2}\right].$
Considering the diagonal space-space components of energy-momentum tensor one can write
\begin{equation*}
 \bar{T^{i}_{j}}\delta_{ij}=-\frac{1}{2a^{2}}\left[\varphi^{\prime2}+\frac{1}{4}H^{2}\varphi^{2}\right]+V,
\end{equation*}
and 
\begin{equation*}
 F^{i}_{j}\delta_{ij}=\frac{1}{4a^{2}}H^{2}\varphi^{2}+\frac{1}{4a^{2}}\left(H\varphi^{2}\right)^{\prime}.
\end{equation*}
From these one can obtain  the expression for pressure as
\begin{equation}
p= \frac{1}{2a^{2}}\left[\varphi^{\prime2}-\frac{1}{4}H^{2}\varphi^{2}\right]-
     \frac{1}{4a^{2}}\left(H\varphi^{2}\right)^{\prime}-V.
\end{equation}
It is easy to notice that the pressure is homogeneous and isotropic. 
All other components of background $T^{\mu}_{\nu}$ are zero. By adding $\varepsilon$ and
$p$ 
\begin{equation}
 \left(\varepsilon+p\right)=\frac{\varphi^{\prime 2}}{a^{2}}+\frac{1}{4a^{2}}H^{2}\varphi^{2}-
                       \frac{1}{4a^{2}}\left(H\varphi^{2}\right)^{\prime}.
\end{equation}
For the instance when the last two terms in the above equations are absent,
one can recover the expression for $(\varepsilon+p)$ of the canonical scalar-field.
Equation of motion for $\varphi$ can be obtained by equating
Equating the divergence of $T^{\mu}_{\nu}$ to zero: 
\begin{equation}
 \varphi^{\prime\prime}+2H\varphi^{\prime}-\frac{3}{4}H^{2}\varphi+V_{,\varphi}=0.
\end{equation}
It should be emphasized that
the above equation for $\varphi$ matches with the equation motion obtained using Euler-Lagrange equation 
as discussed in \cite{bh}.  
However, in the earlier calculations based on non Lorentz invariant model of NSS 
there were mismatches between the equation motions calculated using these 
two methods, e.g. \cite{bh3}.
This is solved because of the additional term $F^{\mu}_{\nu}$ in equation (1.3).
The modified Friedmann equations can be written as 
\begin{eqnarray}
\nonumber H^{2}&=& \frac{1}{1-\tilde{F}}\left[\frac{1}{3M^{2}_{\rm pl}}\left(\frac{\varphi^{\prime 2}}{2}+
                                              a^{2}V\right)\right],\\
 H^{\prime}&=& \frac{1}{1-\tilde{F}}\left[\frac{1}{3M^{2}_{\rm pl}}\left(a^{2}V-\varphi^{\prime 2}\right)+
                          H\tilde{F}^{\prime}\right],
\end{eqnarray}
where, $\tilde{F}=\frac{\varphi^2}{8 M^2_{pl}}$.
One can notice from the above that the condition $\tilde{F} < 1$
is required to be satisfied to ensure the positivity of $H^2$.
Therefore $\varphi$ has to be smaller than $\sqrt{8}M_{\rm pl}$ as mentioned in \cite{bh}. 
We would like to emphasize
that the introduction of $J^{\mu\nu\rho}$ term in equation (1.3) makes the
expressions for $H^2$ and $H^\prime$ different from those obtained in \cite{sh1}.
From what follows we drop the label $cosmo$ on the energy-momentum tensor defined
by equation (1.3).

\section{Perturbed equations}
In this work we closely follow the gauge invariant approach for treating
the cosmological perturbations discussed in Ref. \cite{mukh}.
Total FRLW metric (perturbed + unperturbed) can be written as:
\begin{equation}
 \bar{g}_{\mu\nu}+\delta g_{\mu\nu}=a^{2} \begin{pmatrix}
                 \left(1+2\psi\right) & \mathbb{O}\\
                 \mathbb{O} & \left\{\left(-1+2\phi\right)\delta_{ij}+2h_{ij}\right\}
                 \end{pmatrix}.
\end{equation}
Here $i,j$ denotes space-space components of the metric, $\phi$,$\psi$ are scalar perturbations and 
$h_{ij}$ are traceless and divergence-less tensor perturbations. 
The metric perturbations are functions of space and time. 
We first calculate the perturbations in energy momentum tensor $\delta T^{\mu}_{\nu}$ 
by including the perturbations in $F^{\mu}_{\nu}$ term. Final equations are obtained by 
substituting $\delta T^{\mu}_{\nu}$ into the perturbed Einstein's equations.
Following the structure of unperturbed $\lambda$ and $\dual{\lambda}$ in reference \cite{sh1},
we consider the following form of perturbed $\delta \lambda$ and its dual $\delta\dual{\lambda}$:
\begin{equation*}
\delta\lambda=\frac{\varphi(\eta)}{\sqrt[4]{12}} \begin{pmatrix} 
                                                       -\alpha_{1}e^{i\frac{\pi}{4}}\delta\varphi_{1}\\
                                                        \alpha_{2}\frac{i}{\sqrt{2}}\delta\varphi_{2}\\
                                                        \alpha_{2}\frac{1}{\sqrt{2}}\delta\varphi_{3}\\
                                                        \alpha_{1}e^{i\frac{\pi}{4}}\delta\varphi_{4}
                                                 \end{pmatrix}
                    =\frac{\delta\varphi}{\sqrt[4]{12}} \begin{pmatrix} 
                                                             -\alpha_{1}e^{i\frac{\pi}{4}}\\
                                                              \alpha_{2}\frac{i}{\sqrt{2}}\\
                                                              \alpha_{2}\frac{1}{\sqrt{2}}\\
                                                              \alpha_{1}e^{i\frac{\pi}{4}}
                                                        \end{pmatrix}
\end{equation*}
and
\begin{eqnarray}
\nonumber \delta\dual{\lambda}&=&\frac{\varphi(\eta)}{\sqrt[4]{12}} \begin{pmatrix} 
                                                                     -\alpha_{1}e^{-i\frac{\pi}{4}}\delta\varphi_{1} &
                                                                     -\alpha_{2}\frac{i}{\sqrt{2}}\delta\varphi_{2} &
                                                                      \alpha_{2}\frac{1}{\sqrt{2}}\delta\varphi_{3} &
                                                                      \alpha_{1}e^{-i\frac{\pi}{4}}\delta\varphi_{4}
                                                                    \end{pmatrix}\\
                              &=&\frac{\delta\varphi}{\sqrt[4]{12}} \begin{pmatrix} 
                                                                         -\alpha_{1}e^{-i\frac{\pi}{4}}&
                                                                         -\alpha_{2}\frac{i}{\sqrt{2}}&
                                                                          \alpha_{2}\frac{1}{\sqrt{2}}&
                                                                          \alpha_{1}e^{-i\frac{\pi}{4}}
                                                                    \end{pmatrix},
\end{eqnarray}
with $\delta\varphi_{1}=\delta\varphi_{2}=\delta\varphi_{3}=\delta\varphi_{4}$, $\delta\varphi=\varphi\delta\varphi_{1}$. 
Here $\alpha_{1}=\alpha^{-1}_{2}=\sqrt{\frac{1+\sqrt{3}}{2}}$. $\delta\varphi$ is a function of space and time.
It should be noted that here we have not used the hedgehog ansatz for the unperturbed NSS
like the previous study \cite{sh1}. Instead with the relatively simpler ansatz considered
above one can check that all the equations of the cosmological perturbations given
in \cite{sh1} can be reproduced if the $J^{\mu\nu\rho}$ term is ignored from equation (1.3).

\subsection{Perturbed energy momentum tensors:}
Using equations (3.1, 3.2) we can calculate 
$\delta T^\eta_\eta$, $\delta T^\eta_i$ and $\delta T^i_j (i \neq j)$ components
of the perturbed energy-momentum tensor. 
Below we have enlisted the different components of the energy-momentum
tensor for the scalar perturbations.\\

\noindent
i) \underline{Perturbation of $\varepsilon=T^{\eta}_{\eta}$:}
One can write the general expression for energy as
\begin{equation*}
\varepsilon=X+V,
\end{equation*}
where, $X$ can be written as $X=Y+g_{\eta\eta}\frac{1}{2}\nabla_{\rho}J^{\eta\eta\rho}$, here,
$
Y=\left(\nabla^{\eta} \dual{\lambda}\nabla_{\eta}\lambda\right)-
                 g^{\eta}_{\eta}\left(\frac{1}{2}\nabla_{\mu}\dual{\lambda}\nabla^{\mu}\lambda\right).
$
From the expression of $\varepsilon$ which is a function of X and V we can write
\begin{equation*}
 \delta\varepsilon=\varepsilon_{,X}\delta X+\varepsilon_{,\varphi}\delta\varphi
\end{equation*} and from continuity equation we know 
\begin{equation*}
 \varepsilon^{\prime}=\varepsilon_{,X}X^{\prime}+\varepsilon_{,\varphi}\varphi^{\prime}=
                         -3H\left(\varepsilon+p\right).
\end{equation*}
Eliminating  $\varepsilon_{,\varphi}$ from those two equations we get 
\begin{equation*}
 \delta\varepsilon=\varepsilon_{,X}\left(\delta X-X^{\prime}\frac{\delta\varphi}{\varphi^{\prime}}\right)-
                            3H\left(\varepsilon+p\right)\frac{\delta\varphi}{\varphi^{\prime}}.
\end{equation*}
The perturbation in Y is
\begin{equation*}
 \delta Y=\frac{1}{a^{2}}\left(-\psi\varphi^{\prime2}+\frac{3}{4}\psi H^{2}\varphi^{2}+
                                        \varphi^{\prime}\delta\varphi^{\prime}+
                                        \frac{3}{4}\psi^{\prime}H\varphi^{2}-\frac{3}{4}H^{2}\varphi\delta\varphi\right),
\end{equation*}
while the perturbation in $F^{\eta\eta}$ can be written as
\begin{eqnarray*}
 \delta F^{\eta\eta}= \frac{1}{2}\delta\left(\nabla_{\rho}J^{\eta\eta\rho}\right).
\end{eqnarray*}
Next, the perturbation in the covariant derivative of $J^{\mu\nu\rho}$ can be written as;
\begin{equation*}
 \delta\left(\nabla_{\rho}J^{\eta\eta\rho}\right)=
\partial_{\rho}\delta J^{\eta\eta\rho}+\delta\left(\Gamma^{\eta}_{\sigma\rho}J^{\sigma\eta\rho}+
                        \Gamma^{\eta}_{\sigma\rho}J^{\eta\sigma\rho}+\Gamma^{\rho}_{\sigma\rho}J^{\eta\eta\sigma}\right).
\end{equation*}
Therefore we get after substituting for $\delta\left(\nabla_{\rho}J^{\eta\eta\rho}\right)$
\begin{equation*}
 \delta F^{\eta\eta}=-\frac{1}{4a^{4}}\left(\Delta\psi\right)\varphi^{2}+\frac{3}{2a^{4}}H^{2}\varphi\delta\varphi-
                                       \frac{3}{2a^{4}}\phi^{\prime}H\varphi^{2}-\frac{3}{a^{4}}\psi H^{2}\varphi^{2}.
\end{equation*}
From this one can calculate $\delta X$
\begin{equation*}
 \delta X=-\psi\left(2X\right)+\frac{1}{a^2}\varphi^{\prime}\delta\varphi^{\prime}+
                    \frac{3}{4a^{2}}\left(\psi^{\prime}-2\phi^{\prime}\right)H\varphi^{2}+
               \frac{3}{4a^{2}}H^{2}\varphi\delta\varphi-\frac{1}{4a^{2}}\left(\Delta\psi\right)\varphi^{2},
\end{equation*}
\begin{equation*}
 X^{\prime}=-2HX+\frac{1}{a^{2}}\varphi^{\prime}\varphi^{\prime\prime}+\frac{3}{4a^{2}}HH^{\prime}\varphi^{2}+
                              \frac{3}{4a^{2}}H^{2}\varphi\varphi^{\prime}.
\end{equation*}
Finally one can write  the energy perturbation $\delta \epsilon$ as
\begin{eqnarray}
 \nonumber  \delta\varepsilon&=&\varepsilon_{,X}[2X\left(-\psi+H\frac{\delta\varphi}{\varphi^{\prime}}+
                                         \left(\frac{\delta\varphi}{\varphi^{\prime}}\right)^{\prime}\right)-
                                \frac{3}{4a^{2}}H\varphi^{2}\left(H\frac{\delta\varphi}{\varphi^{\prime}}\right)^{\prime}+
                                \frac{3}{4a^{2}}\left(\psi^{\prime}-2\phi^{\prime}\right)H\varphi^{2}-\\
                   &&          \frac{1}{4a^{2}}\left(\Delta\psi\right)\varphi^{2}]-
                              3H\left(\varepsilon+p\right)\frac{\delta\varphi}{\varphi^{\prime}}.
\end{eqnarray}
\\ \\
ii) \underline{Perturbation of $T^{\eta}_{i}$:}
\begin{equation*}
 \delta T^{\eta}_{i}=\bar{\delta T^{\eta}_{i}}+\delta F^{\eta}_{i}.
\end{equation*}
Now for scalar perturbation 
\begin{equation*}
\bar{\delta T^{\eta}_{i}}=\left[\frac{1}{a^{2}}\varphi^{\prime}\delta\varphi-
                                        \frac{1}{4a^{2}}\left(H\varphi^{2}\right)\psi\right]_{,i},
\end{equation*}
and 
\begin{equation*}
 F^{\eta}_{i}=\left[-\frac{a^{2}}{8}\left(\frac{\psi\varphi^{2}}{a^{4}}\right)^{\prime}-
                        \frac{1}{8a^{2}}H\left(2\varphi\delta\varphi\right)-
                        \frac{1}{4a^{2}}\left(\psi+\phi\right)H\varphi^{2}+
                        \frac{1}{8a^{2}}\phi^{\prime}\varphi^{2}\right]_{,i}.
\end{equation*}
And we get 
\begin{eqnarray}
 \nonumber \delta T^{\eta}_{i}&=&\left[\frac{1}{a^{2}}\varphi^{\prime}\delta\varphi-
                                                     \frac{1}{4a^{2}}\left(H\varphi^{2}\right)\psi\right]_{,i}+\\
                     &&\left[-\frac{a^{2}}{8}\left(\frac{\psi\varphi^{2}}{a^{4}}\right)^{\prime}-
                       \frac{1}{8a^{2}}H\left(2\varphi\delta\varphi\right)-
                     \frac{1}{4a^{2}}\left(\psi+\phi\right)H\varphi^{2}+
            \frac{1}{8a^{2}}\phi^{\prime}\varphi^{2}\right]_{,i}.
\end{eqnarray}
\\
iii) \underline{Perturbation of $T^{i}_{j}\left(i\neq j\right)$:}
\begin{equation*}
 \delta T^{i}_{j}=\delta\bar{T^{i}_{j}}+F^{i}_{j}.
\end{equation*}
Now for scalar perturbation, 
$\delta\bar{T^{i}_{j}}=0$ and 
$F^{i}_{j}=-\frac{1}{4a^{2}}\left(\partial_{i}\partial_{j}\phi\right)\varphi^{2}$ for $i\neq j$. 
Therefore
\begin{equation}
 \delta T^{i}_{j}=-\frac{1}{4a^{2}}\left(\partial_{i}\partial_{j}\phi\right)\varphi^{2}  \qquad
\left(i\neq j\right).
\end{equation}
\\
\subsection{Perturbed Einstein's Equation:}
Perturbed Einstein's equation can be written as:
$$
\delta G^{\mu}_{\nu}=8\pi G \delta T^{\mu}_{\nu},
$$
where $\delta G^{\mu}_{\nu}$ is the perturbed Einstein's tensor.
The scalar part of perturbed Einstein's equations are given below,
\begin{eqnarray*}
\Delta\phi-3H\left(\phi^{\prime}+H\psi\right)= 4 \pi G a^{2}\delta T^{\eta}_{\eta}
\end{eqnarray*}
\begin{eqnarray*}
-\left[2\phi^{\prime\prime}+2\frac{a^{\prime}}{a}\left(\psi^{\prime}+2\phi^{\prime}\right)-
2\left\{\left(\frac{a^{\prime}}{a}\right)^{2}-2\frac{a^{\prime\prime}}{a}\right\}\psi+\Delta\left(\psi-\phi\right)\right]\delta_{ij}+
\partial_{i}\partial_{j}\left(\psi-\phi\right) = 8\pi G a^{2}\delta T^{i}_{j}
\end{eqnarray*}
\begin{eqnarray}
\left(\phi^{\prime}+H\psi\right)_{,i}=4\pi G a^{2}\delta T^{\eta}_{i},
\end{eqnarray}
In the previous sub-section we have already calculated 
the scalar perturbations for the various components of the energy-momentum tensor.
The tensor part of the perturbed Einstein's equations can be written as,
\begin{equation*}
 h^{\prime\prime}_{ij}+2Hh^{\prime}_{ij}-\Delta h_{ij}=-16\pi Ga^{2}\delta T^{i}_{j(T)},
\end{equation*}
where subscript $T$ on the energy-momentum tensor denotes the tensor part. 
Next, consider the space-space components of the Einstein equation with $(i\neq j)$.\\ \\
\noindent
i) \underline{$\delta G^{i}_{j}=8\pi G\delta T^{i}_{j}$:}
Using the expression of $\delta T^{i}_{j}$ when $i\neq j$ from equation (3.5) one can write,
\begin{equation*}
     \partial_{i}\partial_{j}\left(\psi-\phi\right)=\partial_{i}\partial_{j}\left(-2\tilde{F}\phi\right),
\end{equation*}
where $\tilde{F}=\pi G\varphi^{2}=\frac{\varphi^{2}}{8M^{2}_{\rm PL}}$. 
In the case of the standard inflation driven by a canonical scalar-field,
$\delta{T^{i}_{j}}=0$ for $i\neq j$ and $\phi=\psi$. However, this is
no longer true for a NSS driven inflation. The above equation implies
that  the condition $\psi=(1-2\tilde{F})\phi$ needs to be satisfied by the metric
and the NSS perturbations. 
In the previous study  using a NSS field
\cite{sh1}, $\delta\bar{T^{i}_{j}}=0$ for $i\neq j$. That's why in \cite{sh1} we got 
$\phi=\phi$.
Here inequality between  $\psi$ and $\phi$ arises because of the extra
$F^{\mu\nu}$ term in the energy-momentum tensor. 
We consider $\tilde{F}$ to be a very small quantity and from here onwards 
we will write the equations up to 
the linear order in $\tilde{F}$. \\

\noindent
ii) \underline{$\delta G^{\eta}_{i}=8\pi G\delta T^{\eta}_{i}$:} Using the last equation of (3.6) we get
\begin{eqnarray*}
 \frac{2}{a^{2}}\left(\phi^{\prime}+H\psi\right)_{,i}
      &=& 8\pi G[\frac{1}{a^{2}}\varphi^{\prime}\delta\varphi-\frac{1}{4a^{2}}\left(H\varphi^{2}\right)\psi
          -\frac{a^{2}}{8}\left(\frac{\psi\varphi^{2}}{a^{4}}\right)^{\prime}-
           \frac{1}{8a^{2}}H\left(\delta\dual{\lambda}\lambda+\dual{\lambda}\delta\lambda\right)-\\
&&  \frac{1}{4a^{2}}\left(\psi+\phi\right)H\varphi^{2}+
      \frac{1}{8a^{2}}\phi^{\prime}\varphi^{2}]_{,i},
\end{eqnarray*}
or,
\begin{eqnarray*}
 \left(\phi^{\prime}+H\psi\right)=
                 4\pi Ga^{2}\left(\frac{\varphi^{\prime 2}}{a^{2}}\right)\frac{\delta\varphi}{\varphi^{\prime}}-H\tilde{F}\phi-
                 \left(\frac{\psi^{\prime}-\phi^{\prime}}{2}\right)\tilde{F}-
                 \frac{\tilde{F}^{\prime}}{2}\left(H\frac{\delta\varphi}{\varphi^{\prime}}+\psi\right).
\end{eqnarray*}
Substituting $\psi=(1-2\tilde{F})\phi$ in the right hand side of the above equation,
\begin{equation}
 \left(\phi^{\prime}+H\psi\right) \simeq 
                4\pi Ga^{2}\left(\varepsilon+p\right)\frac{\delta\varphi}{\varphi^{\prime}}+
                \left[\left(H\tilde{F}\right)^{\prime}-H^{2}\tilde{F}-
                 \frac{H\tilde{F}^{\prime}}{2}\right]\frac{\delta\varphi}{\varphi^{\prime}}-
                \left(H\tilde{F}+\frac{\tilde{F}^{\prime}}{2}\right)\phi.
\end{equation}
Again setting $\psi=(1-2\tilde{F})\phi$ in the left hand side and multiplying both sides by $\frac{a^{2}}{H}$ 
one may obtain,
\begin{eqnarray}
\nonumber \left(\frac{a^{2}}{H}\phi\right)^{\prime} &\simeq& 
 \left[\frac{4\pi Ga^{4}}{H^{2}}\left(\varepsilon+p\right)+
                                    \frac{a^{2}}{H}\left\{\frac{\left(H\tilde{F}\right)^{\prime}}{H}-H\tilde{F}-
                                    \frac{\tilde{F}^{\prime}}{2}\right\}\right]\left(H\frac{\delta\varphi}{\varphi^{\prime}}+\phi\right)+\\
                             &&       \left(2H\tilde{F}-\frac{\left(H\tilde{F}\right)^{\prime}}{H}\right)\frac{a^{2}\phi}{H}.
\end{eqnarray}
\\ \\
iii) \underline{$\delta G^{\eta}_{\eta}=8\pi G\delta T^{\eta}_{\eta}=8\pi G\delta\delta\varepsilon$:} Now the first equation in (3.6)
implies 
\begin{equation*}
 \Delta\phi-3H\left(\phi^{\prime}+H\psi\right)=4\pi Ga^{2}\delta\varepsilon.
\end{equation*}
Using the expression of $\left(\phi^{\prime}+H\psi\right)$ from equation (3.7) we get
\begin{equation*}
 \Delta\phi-3H\left[4\pi Ga^{2}\left(\varepsilon+p\right)\frac{\delta\varphi}{\varphi^{\prime}}+
   \left\{\left(H\tilde{F}\right)^{\prime}-H^{2}\tilde{F}-
   \frac{H\tilde{F}^{\prime}}{2}\right\}\frac{\delta\varphi}{\varphi^{\prime}}-
   \left(H\tilde{F}+\frac{\tilde{F}^{\prime}}{2}\right)\phi\right]
   \simeq 4\pi Ga^{2}\delta\varepsilon.
\end{equation*}
Similarly, using the expression of $\psi$ from equation (3.7) in the expression of $\delta\varepsilon$ we get,
\begin{eqnarray*}
 \delta\varepsilon  &\simeq&  
       \varepsilon_{,X}\frac{2X}{H}\left[\left(H\frac{\delta\varphi}{\varphi^{\prime}}+\phi\right)^{\prime}-\left\{(H\tilde{F})^{\prime}-
       H^{2}\tilde{F}-\frac{H\tilde{F}^{\prime}}{2}\right\}\frac{\delta\varphi}{\varphi^{\prime}}+
                                            \left(H\tilde{F}+\frac{\tilde{F}^{\prime}}{2}\right)\phi\right]-\\
    &&  \varepsilon_{,X}\frac{3}{4a^{2}}H\varphi^{2}\left(H\frac{\delta\varphi}{\varphi^{\prime}}+\phi\right)^{\prime}+
          \varepsilon_{,X}\frac{3}{4a^{2}}\left(\psi^{\prime}-\phi^{\prime}\right)H\varphi^{2}-
          \varepsilon_{,X}\frac{1}{4a^{2}}\left(\Delta\psi\right)\varphi^{2}-\\
    && 3H\left(\varepsilon+p\right)\frac{\delta\varphi}{\varphi^{\prime}}.
\end{eqnarray*}
Finally using $\psi=(1-2\tilde{F})\phi$ in the above expression of $\delta\varepsilon$, up to linear order in 
$\tilde{F}$ the Einstein's equation becomes,

\begin{eqnarray}
\nonumber \left(1+\varepsilon_{,X}\tilde{F}\right)\Delta\phi  &\simeq&  
                       \left(4\pi Ga^{2}\varepsilon_{,X}\frac{2X}{H}-
                       \varepsilon_{,X}3H\tilde{F}\right)\left(H\frac{\delta\varphi}{\varphi^{\prime}}+\phi\right)^{\prime}+\\
 \nonumber            &&   \left(3H-4\pi Ga^{2}\varepsilon_{,X}\frac{2X}{H}\right)\left\{\frac{\left(H\tilde{F}\right)^{\prime}}{H}-
                           H\tilde{F}-\frac{\tilde{F}^{\prime}}{2}\right\}\left(H\frac{\delta\varphi}{\varphi^{\prime}}+\phi\right)-\\
                      &&       \left(3H-4\pi Ga^{2}\varepsilon_{,X}\frac{2X}{H}\right)\frac{\left(H\tilde{F}\right)^{\prime}}{H}\phi.
\end{eqnarray}
In order to calculate the power spectrum for $\phi$ and $\delta\varphi$, 
we have to solve equations (3.8,3.9). These equations are highly coupled
and one may need to decouple them. For simplicity we first write equations (3.8,3.9) in a
different notations as below:
\begin{eqnarray}
  x^{\prime}&=&A_{1}y+B_{1}x,\\
              A_{2}\Delta x&=&B_{2}y^{\prime}+C_{2}y-D_{2}x.
\end{eqnarray}
where, 
\begin{eqnarray*}
       x &=& \left(\frac{a^{2}\phi}{H}\right),\\
       y &=&\left(H\frac{\delta\varphi}{\varphi^{\prime}}+\phi\right),\\
A_{1} &=& \frac{4\pi Ga^{4}}{H^{2}}\left(\varepsilon+p\right)+\frac{a^{2}}{H}\left[\frac{(H\tilde{F})^{\prime}}{H}-H\tilde{F}-
                                    \frac{\tilde{F}^{\prime}}{2}\right],\\
B_{1} &=& \left(2H\tilde{F}-\frac{\left(H\tilde{F}\right)^{\prime}}{H}\right),\\
A_{2} &=& \left(1+\varepsilon_{,X}\tilde{F}\right),
\end{eqnarray*}
\begin{eqnarray*}
B_{2} &=& \frac{a^{2}}{H}\left(4\pi Ga^{2}\varepsilon_{,X}\frac{2X}{H}-\varepsilon_{,X}3H\tilde{F}\right),\\
C_{2} &=& \frac{a^{2}}{H}\left(3H-4\pi Ga^{2}\varepsilon_{,X}\frac{2X}{H}\right)
                  \left[\frac{\left(H\tilde{F}\right)^{\prime}}{H}-H\tilde{F}-\frac{\tilde{F}^{\prime}}{2}\right],\\
D_{2} &=& \left(3H-4\pi Ga^{2}\varepsilon_{,X}\frac{2X}{H}\right)\frac{\left(H\tilde{F}\right)^{\prime}}{H}.
\end{eqnarray*}
 $y$ can be eliminated from equation (3.11) by using equation (3.10) and
the decoupled equation for $x$ can be written as,
\begin{equation}
x^{\prime\prime}-
            \frac{A_{1}A_{2}}{B_{2}}\Delta x+\left[A_{1}\left\{\left(\frac{1}{A_{1}}\right)^{\prime}-\frac{B_{1}}{A_{1}}\right\}+
            \frac{C_{2}}{B_{2}}\right]x^{\prime}-
            \left[A_{1}\left(\frac{B_{1}}{A_{1}}\right)^{\prime}+C_{2}\frac{B_{1}}{B_{2}}+D_{2}\frac{A_{1}}{B_{2}}\right]x=0.
\end{equation}
\noindent
Next,  it is useful to substitute $x=u(\eta,\vec{x})f(\eta)$ in the above equation
\begin{eqnarray*}
\nonumber &&  u^{\prime\prime}-\frac{A_{1}A_{2}}{B_{2}}\Delta u+
               \left[2\frac{f^{\prime}}{f}+\left\{A_{1}\left(\frac{1}{A_{1}}\right)^{\prime}-B_{1}+
               \frac{C_{2}}{B_{2}}\right\}\right]u^{\prime}+\\
     &&  \left[\frac{f^{\prime\prime}}{f}+\left\{A_{1}\left(\frac{1}{A_{1}}\right)^{\prime}-B_{1}+
           \frac{C_{2}}{B_{2}}\right\}\frac{f^{\prime}}{f}-
           \left\{A_{1}\left(\frac{B_{1}}{A_{1}}\right)^{\prime}+C_{2}\frac{B_{1}}{B_{2}}+
           D_{2}\frac{A_{1}}{B_{2}}\right\}\right]u=0.
\end{eqnarray*}
By equating the coefficient of $u^{\prime}$  to zero one can gets 
\begin{eqnarray*}
 f &=& \rm exp\left[-\frac{1}{2}\int\left\{A_{1}\left(\frac{1}{A_{1}}\right)^{\prime}-B_{1}+\frac{C_{2}}{B_{2}}\right\}d\eta\right]\\
   &=& \sqrt{A_{1}}\rm exp\left[\frac{1}{2}\int\left(B_{1}-\frac{C_{2}}{B_{2}}\right)d\eta\right]
\end{eqnarray*}
Here ${f}^\prime$ and $f^{\prime\prime}$ can be written as
\begin{eqnarray*}
 \frac{f^{\prime}}{f}           &=& -\frac{1}{2}\left[A_{1}\left(\frac{1}{A_{1}}\right)^{\prime}-
                                                                            B_{1}+\frac{C_{2}}{B_{2}}\right]\\
\frac{f^{\prime\prime}}{f}   &=& \left[-\frac{1}{2}\left\{A_{1}\left(\frac{1}{A_{1}}\right)^{\prime}-
                                                          B_{1}+\frac{C_{2}}{B_{2}}\right\}\right]^{2}-
                                   \frac{1}{2}\left[A_{1}\left(\frac{1}{A_{1}}\right)^{\prime}-B_{1}+
                                   \frac{C_{2}}{B_{2}}\right]^{\prime}.
\end{eqnarray*}
Finally the generalized Mukhanov-Sasaki equation can be written as
\begin{eqnarray*}
 \nonumber &&  u^{\prime\prime}-\frac{A_{1}A_{2}}{B_{2}}\Delta u+
                         \left[-\frac{1}{4}\left\{A_{1}\left(\frac{1}{A_{1}}\right)^{\prime}-B_{1}+\frac{C_{2}}{B_{2}}\right\}^{2}-
                         \frac{1}{2}\left\{A_{1}\left(\frac{1}{A_{1}}\right)^{\prime}-B_{1}+\frac{C_{2}}{B_{2}}\right\}^{\prime}-\right.\\
                  && \left. \left\{A_{1}\left(\frac{B_{1}}{A_{1}}\right)^{\prime}+C_{2}\frac{B_{1}}{B_{2}}+
                         D_{2}\frac{A_{1}}{B_{2}}\right\}\right]u=0,
\end{eqnarray*}
which one may  write in a more simplified form as
\begin{equation}
 u^{\prime\prime}+\left(1+A\right)k^{2}u-\left(\frac{\theta^{\prime\prime}}{\theta}+B\right)u=0,
\end{equation}
Here both $A$ and $B$ are functions of $\tilde{F}$ and its derivatives. In the limit $\tilde{F}\rightarrow 0$
both $A, B \rightarrow 0$ and one recovers the standard Mukhanov-Sasaki equation for
a canonical scalar-field \cite{mukh}. 
\par
The coefficient of the $k^2$ term in equation (3.13) can be regarded as the square of 
sound speed ($c^2_s$), which implies $c^2_s=\left(1+A\right)=\frac{A_{1}A_{2}}{B_{2}}$. 
Using the expressions of $A_{1}$,$A_{2}$ and $B_{2}$ in terms of 
background quantities, we can write $c^{2}_{s}$ after some algebra :
\begin{equation}
 c^{2}_{s}\simeq 1+\tilde{F}\left[1-\frac{1}{3}\frac{\tilde{F}^{\prime}}{H\tilde{F}}
                            \frac{1}{\left(1+\frac{p}{\varepsilon}\right)_{\rm can}}\right]
\end{equation}
where, $\left(1+\frac{p}{\varepsilon}\right)_{\rm can}$ can be obtained by 
setting $\tilde{F}$ terms to zero in equations (2.4-2.5).
On galactic scale $\frac{1}{\left(1+\frac{p}{\varepsilon}\right)_{\rm can}}\sim10^{-2}$ 
(for example one can see Ref. \cite{mukh}) . Again, in slow-roll 
inflation we can consider $\frac{\tilde{F}^{\prime}}{H\tilde{F}}\ll1$. 
 Thus one can write $c^{2}_{s}\simeq 1+\tilde{F}.$

\section{Calculation of power spectrum}
From the solutions of equation (3.13) the power spectrum for the scalar-perturbations
can be calculated. In what follows we closely follow the method of the power-spectrum
calculations given in \cite{mukh} for a canonical scalar-field.\\

\noindent
i) \underline{Short wavelength(large $k$) region :} For a short wavelength regime (or large $k$),
  we can neglect 
$\left(\frac{\theta^{\prime\prime}}{\theta}+B\right)$ term with respect to $\left(1+A\right)k^{2}$ term in equation (3.13)
and write 
\begin{equation}
 u^{\prime\prime}+\left(1+A\right)k^{2}u=0.
\end{equation}
One may look for the solution of equation (4.1) in the form 
$u=c\left(\eta\right)exp\left[ik\int\sqrt{1+A}d\eta\right]$.
Substituting this back into  equation(4.1) we get a 2nd order equation for $c(\eta)$,
\begin{equation}
 c^{\prime\prime}+ikc^{\prime}\left(1+\frac{A}{2}\right)+ikc\frac{A^{\prime}}{2}=0,
\end{equation}
where we have considered $A$ to be a small quantity and write $\sqrt{1+A}\simeq1+\frac{A}{2}$. 
Next, We look for an approximate solution of equation(4.2) by regarding $A$ and $A^{\prime}$ to be small. 
Thus we consider $c\approx c_{0}+\bar{c}$ with $|c_0|> |\bar{c}|$
and $\bar{c}$ is of the same order of $A$ and $A^{\prime}$. Equations for $c_0$ and $\bar{c}$ 
can be written as follows,
\begin{eqnarray}
 \nonumber && c_{0}^{\prime\prime}+ikc_{0}^{\prime}=0\\
               && \bar{c}^{\prime\prime}+ikc_{0}^{\prime}\frac{A}{2}+ik\bar{c}^{\prime}+ikc_{0}\frac{A^{\prime}}{2}=0.
\end{eqnarray}
 The solution for $c_{0}$  can be written as 
\begin{equation*}
 c_{0}=b_{2}-\frac{b_{1}e^{-ik\eta}}{ik},
\end{equation*}
where $b_{1}$ and $b_{2}$ are the constants of integration. Solution for
$\bar{c}$ can be obtained as
\begin{equation*}
 \bar{c}={e}^{-ik\eta}\int\left(b_{1}-ikb_{2}e^{ik\eta}\right)\frac{A}{2}d\eta.
\end{equation*}
Finally we get,
\begin{equation*}
c(\eta)=b_{2}-\frac{b_{1}
e^{-ik\eta}}{ik}+e^{-ik\eta}\int\left(b_{1}-ikb_{2}e^{ik\eta}\right)\frac{A}{2}d\eta.
\end{equation*}
Since in the limit when $A=0$ one should get the solution of the canonical scalar-field \cite{mukh},
we set $b_{1}=0$ and $b_{2}=-\frac{i}{k^{\frac{3}{2}}}$. Thus one can write solution of equation (4.1) as
\begin{equation}
u=-\frac{i}{k^{\frac{3}{2}}}\left[1-ike^{-ik\eta}\int e^{ik\eta}\frac{A}{2}d\eta\right]exp\left[ ik\int\sqrt{1+A}d\eta\right].
\end{equation}
Finally one can obtain
\begin{eqnarray}
\nonumber \phi&=&-\frac{i}{k^{\frac{3}{2}}}\left[1-ike^{-ik\eta}\int e^{ik\eta}\frac{A}{2}d\eta\right]
                 \left\{\frac{H}{a^{2}}\sqrt{A_{1}}exp\left[\frac{1}{2}\int\left(B_{1}-\frac{C_{2}}{B_{2}}\right)d\eta\right]\right\}\times\\
                                                &&  exp\left[ ik\int\sqrt{1+A}d\eta\right].
\end{eqnarray}
From this the power spectrum for $\phi$ in case of large $k$(small wavelength) can be found to be
\begin{eqnarray}
\nonumber \delta^{2}_{\phi} &=& |\phi|^{2}k^{3}\\
                            &=& \left\{\frac{H^{2}}{a^{4}}A_{1}exp\left[\int\left(B_{1}-\frac{C_{2}}{B_{2}}\right)d\eta\right]\right\}
                                \left[1-ike^{-ik\eta}\int e^{ik\eta}\frac{A}{2}d\eta\right]^{2}.
\end{eqnarray}
The at large $k$ (or small wavelength) scales the power spectrum of $\phi$ is not
a scale-invariant. However it can become a scale invariant if $A$ can be regarded
as a constant. \\ \\
\noindent
ii) \underline{Large wavelength(Small $k$) region :} In case of small $k$ regime one can neglect
 $(1+A)k^{2}$ term with respect to 
$(\frac{\theta^{\prime\prime}}{\theta}+B)$ term. In this case we write the equation (3.13)
can be written as
\begin{equation}
u^{\prime\prime}- \left(\frac{\theta^{\prime\prime}}{\theta}+B\right)u=0.
\end{equation}
It is useful to look for the solution of $u$ in the form $u=u_{can}g$ where, 
$u_{can}$ is the solution when $B=0$ i.e. no effect of non standard spinor is considered. This implies that 
in the $B\rightarrow 0$ limit 
$g=1$. Now substituting for $u$ into equation(4.7) we get the equation for $g$ as
\begin{equation}
 g^{\prime\prime}+2\left(\frac{u_{can}^{\prime}}{u_{can}}\right)g^{\prime}-Bg=0.
\end{equation}
For the case when $B=B(\tilde{F})$ is a small quantity, an approximate solution of
$g\approx (g_0+\bar{g})$ with $|g_0|>|\bar{g}|$ can be obtained in a manner similar to
that discussed in the previous section.
From equation (4.8) we get
\begin{eqnarray}
\nonumber && g_{0}^{\prime\prime}+2\left(\frac{u^{\prime}_{can}}{u_{can}}\right)g_{0}^{\prime}=0\\
                 && \bar g^{\prime\prime}+2\left(\frac{u^{\prime}_{can}}{u_{can}}\right)\bar g^{\prime}=Bg_{0}.
\end{eqnarray}
From the equation for $g_{0}$ we get
\begin{equation*}
 g_{0}=c_{1}+\int\left(\frac{c_{2}}{u^{2}_{can}}\right)d\eta,
\end{equation*}
where $c_{1}$ and $c_{2}$ are constants of integration. 
Plugging this solution of $g_{0}$ into the equation for 
$\bar{g}$ and solving the
inhomogeneous equation, we can write get  $\bar{g}$ as
\begin{equation*}
 \bar{g}=\int\frac{1}{u^{2}_{can}}\left[\int Bu^{2}_{can}d\eta\right]d\eta.
\end{equation*}
Therefore we get, 
\begin{equation*}
g=c_{1}+\int\left(\frac{c_{2}}{u^{2}_{can}}\right)d\eta+
               \int\frac{1}{u^{2}_{can}}\left[\int Bu^{2}_{can}d\eta\right]d\eta.
\end{equation*}
Since $g=1$ when $B=0$, one can set  $c_{1}=1$ and $c_{2}=0$. 
The approximate solution for $u$ can be written as
\begin{equation}
u \simeq u_{can}\left(1+\int\frac{1}{u^{2}_{can}}\left[\int Bu^{2}_{can}d\eta\right]d\eta\right).
\end{equation}
Thus one can notice from the expression of $u$ in the long wavelength(small $k$) regime
that the resultant power spectrum is a scale invariant quantity.
Therefore in a long wavelength regime we can write
\begin{equation*}
 \phi \simeq \frac{H}{a^{2}}\sqrt{A_{1}}exp\left[\frac{1}{2}\int\left(B_{1}-\frac{C_{2}}{B_{2}}\right)d\eta\right]
                       u_{can}\left(1+\int\frac{1}{u^{2}_{can}}\left[\int Bu^{2}_{can}d\eta\right]d\eta\right)
\end{equation*}
Finally we get power spectrum of $\phi$ as
\begin{eqnarray}
\nonumber \delta^{2}_{\phi} 
             &=& |\phi|^{2}k^{3}\\
\nonumber             &=& \delta^{2}_{\phi(can)}\left[1-\frac{H\tilde{F^{\prime}}}{8\pi Ga^{2}\left(\varepsilon+p\right)_{can}}\right]
               exp\left[\int\left(B_{1}-\frac{C_{2}}{B_{2}}\right)d\eta\right]\times\\
              &&\left(1+\int\frac{1}{u^{2}_{can}}\left[\int Bu^{2}_{can}d\eta\right]d\eta\right)^{2}
\end{eqnarray}
Now as $\int\frac{1}{u^{2}_{can}}\left[\int Bu^{2}_{can}d\eta\right]d\eta$ are $k$ independent, we get that power spectrum 
of $\phi$ for large wavelength(small $k$) is scale independent. Taking logarithm on both side we get 
\begin{eqnarray*}
 \ln\delta^{2}_{\phi}&=&\ln\delta^{2}_{\phi(can)}+
                                      \ln\left[1-\frac{H\tilde{F^{\prime}}}{8\pi Ga^{2}\left(\varepsilon+p\right)_{can}}\right]+
                                      \left[\int\left(B_{1}-\frac{C_{2}}{B_{2}}\right)d\eta\right]+\\
                               && 2\ln\left(1+\int\frac{1}{u^{2}_{can}}\left[\int Bu^{2}_{can}d\eta\right]d\eta\right).
\end{eqnarray*}
Spectral index for scalar perturbation can be written as 
\begin{equation*}
 n_{s}-1=\frac{d\ln\left(\delta^{2}_{\phi}\right)}{d\ln k}.
\end{equation*}
At the time of horizon crossing $(c_{s}k=aH)$, derivative with respect to $\ln k$ can be approximated as 
$d\ln k=\frac{1}{H}d\eta$ (here we have considered that variation of sound velocity and Hubble parameter are very 
small, therefore can be neglected). 
Therefore in the expression for the spectral index all the logarithmic derivatives can  be
replaced with time derivatives and finally we get
\begin{eqnarray}
\nonumber n_{s}-1 &=&\frac{1}{H}\left(\ln\delta^{2}_{\phi(can)}\right)^{\prime}+
              \frac{1}{H}\left(\ln\left[1-\frac{H\tilde{F^{\prime}}}{8\pi Ga^{2}\left(\varepsilon+p\right)_{can}}\right]\right)^{\prime}+
              \frac{1}{H}\left(B_{1}-\frac{C_{2}}{B_{2}}\right)+\\
              &&\frac{2}{H}\left[\ln\left(1+\int\frac{1}{u^{2}_{can}}\left[\int Bu^{2}_{can}d\eta\right]d\eta\right)\right]^{\prime}.
\end{eqnarray}
In the case of a slow roll if $A$ is a measurable quantity then $\frac{A^{\prime}}{HA}$ is very small and it can be 
neglected. So we argue that in the above expression we can neglect the second and last term. In the case of a
canonical scalar-field we can write the first term in equation (4.12) following Ref.\cite{mukh} 
as 
\begin{equation}
 \frac{1}{H}\left(\ln\delta^{2}_{\phi(can)}\right)^{\prime} \simeq -3\left(1+\frac{p}{\varepsilon}\right)_{\rm can}.
\end{equation}
But in the case of NSS the correction term $\frac{1}{H}(B_{1}-\frac{C_{2}}{B_{2}})$ can be approximated as 
\begin{equation*} 
 \frac{1}{H}\left(B_{1}-\frac{C_{2}}{B_{2}}\right) \simeq \frac{3H^{2}}{4\pi G\varphi^{\prime 2}}\tilde{F}.
\end{equation*}
Using the Friedmann's equation and keeping the terms up to linear order in $\tilde F$ we write
\begin{equation}
 \frac{1}{H}\left(B_{1}-\frac{C_{2}}{B_{2}}\right) \simeq 
                 2\frac{1}{\left(1+\frac{p}{\varepsilon}\right)_{\rm can}}\tilde{F}.
\end{equation}
Finally using (4.13) and (4.14) we get spectral index for scalar perturbation as 
\begin{equation}
 n_{s}=1-3\left(1+\frac{p}{\varepsilon}\right)_{\rm can}+
               2\frac{1}{\left(1+\frac{p}{\varepsilon}\right)_{\rm can}}\tilde{F}.
\end{equation}
On galactic scale the canonical terms $\left(1+\frac{p}{\varepsilon}\right)_{\rm can}$  
can be estimated as $10^{-2}$ and $\varepsilon_{\rm can}$ can be estimated as $10^{-12}$
of the Planckian density\cite{mukh}. Then equation (4.15) can be written as
\begin{equation}
 n_{s}-1=-0.03+200\tilde{F}.
\end{equation}
WMAP 7 years data suggests $n_{s}=0.968\pm 0.012$ with 68 \% CL \cite{wmap7}.
Therefore from equation (4.16)  we can understand 
that, to get $n_{s}$ closer to the observed value, $\tilde{F}$ has to be smaller than $10^{-4}$.
$\tilde{F}$ is the only new feature which NSS driven inflation brings over the inflationary 
scenario driven by canonical scalar field. Although $\tilde{F}$ is not a part of potential in the theory, its
value may be estimated from $V$.
As the potential $V(\varphi)$ is the dominant term in $\varepsilon_{\rm can}$, we can write
$\varepsilon_{\rm can}/\varepsilon_{\rm PL}\sim \frac{V}{\varepsilon_{\rm PL}}\sim
\frac{V}{M^{4}_{\rm PL}}\sim 10^{-12}$. 
Now from different models of potentials we can estimate $\tilde{F}$. For example, if we consider
$\varphi^{4}$ kind of potential then $\varepsilon_{\rm can}/\varepsilon_{\rm PL}$ 
becomes $\tilde{F}^{2}$ and from the value of $\varepsilon_{\rm can}$ we can estimate 
$\tilde{F}\sim10^{-6}$ which is consistent with the NSS model. Upcoming experiments like PLANCK\cite{planck}
can further constrain $\tilde{F}$ by measuring $n_{s}$ more accurately. \\
In the case of a canonical scalar-field inflation it is well known that at a large scale the power-spectrum 
of tensor perturbation is
\cite{mukh} 
 $\delta^{2}_{\rm h(can)}\simeq \frac{8}{\pi}H^{2}$.
But for the present case the power-spectrum for the tensor perturbation is modified to
\begin{equation}
 \delta^{2}_{\rm h}\simeq \frac{8}{\pi}H^{2}\times f\left(\tilde{F}\right).
\end{equation}
Thus when $\tilde{F}\rightarrow 0$, $f\left(\tilde{F}\right)\rightarrow 1$ and we get the power-spectrum
of the tensor perturbations for a canonical scalar-field. Since $\tilde{F}$ is a small
quantity, the tensor to scalar ratio of the power spectrum for a NSS still be very small.


In conclusion,we have studied the cosmological perturbations generated by the
inflation driven by a Lorentz invariant NSS model. We find that the
the usual condition for the gravitational potentials $\phi$ and $\psi$ for
scalar-perturbations i.e. $\delta{T}^i_{j}=0$ giving $\psi=\phi$ is
modified to $\psi=\left(1-2\tilde{F}\right)\phi$. 
We have also shown that the perturbations are nearly scale invariant and
the hedgehog ansatz is not required.
More importantly we have 
calculated the power-spectrum and spectral index for the metric perturbation. 
The model predicts the running spectral index which allows for a wide
range of $\tilde{F}$.
For the case  $\tilde{F}=0$  
one gets back the expressions for the power spectrum and spectral
index for a canonical scalar-field. 
Further our analysis shows that the calculated value of the
spectral index $n_s$ can match to the value obtained from the WMAP data
if there is an upper bound on the parameter $\tilde{F}<10^{-4}$.
Our analysis shows that the sound speed of the perturbation is not a constant
but dependent on time.
However, the expression of $c^{2}_{s}\simeq 1+\tilde{F}$ and the upper
bound on $\tilde{F}$ imply that $c^2_s\sim 1$.
Finally the tensor to scalar ratio of the power spectrum remains much
smaller as in the case of a scalar-field inflation due the upper bound on
$\tilde{F}$

\begin{acknowledgments} We would like to thank Suratna Das, V. Mukhanov and S. Shankaranarayanan for useful discussions.
\end{acknowledgments}

\end{document}